\documentclass[preprint,aps]{revtex4}
\usepackage{graphicx}
\usepackage{bm}

\begin{document}
\title{Massive scalar field instability in Kerr spacetime}
\author{Matthew J. Strafuss and Gaurav Khanna}
\affiliation{285 Old Westport Road, Physics Department,\\
University of Massachusetts at Dartmouth,\\
Dartmouth MA 02747.}

\date{\today}

\begin{abstract}
We study the Klein-Gordon equation for a massive scalar field in Kerr spacetime in the time-domain. We demonstrate that under conditions of super-radiance, the scalar field becomes unstable and its amplitude grows without bound. We also estimate the growth rate of this instability.  
\end{abstract}

\maketitle

\section{Introduction}
Various studies of the dynamics of scalar fields in black hole spacetimes have been done by many researchers in literature. Most of these that relate to super-radiance and unstable behavior have been performed in the frequency-domain \cite{detweiler}\cite{old}, while time-domain based work has only recently picked up momentum \cite{pablo} \cite{pablo-scalar} \cite{massive}.  

Consider the scattering of a wave (massless scalar, electromagnetic or gravitational) off a Kerr black hole. Generically, we expect that the incident wave would be partly reflected by the hole and propagate outward while the remaining would go down into the hole. However, under conditions of super-radiance, the reflected wave emerges amplified compared to the original incident wave! In this case of super-radiant scattering, energy and angular momentum flow out of the hole and it has been proposed \cite{press} to use this effect to create a ``black hole bomb''. To create a bomb, one places ``mirrors''  surrounding the Kerr hole, and they cause the amplified scattered outgoing wave to simply get reflected back to the hole  which causes amplification again. This process continues on and we ultimately obtain an unbounded growth in the amplitude of the wave. In the next section of this article we demonstrate this effect using a massless scalar and also an electromagnetic wave packet in Kerr spacetime using time-domain black hole perturbation theory (Teukolsky equation \cite{teuk}).

It has also been shown that if one uses a massive scalar field instead, the mass of the field itself provides a natural mirror \cite{detweiler}\cite{damour}. The main reason for this can be seen by examining the form of the potential function for large $r$. If we write the Klein-Gordon equation in a black hole background in terms of a function that scales the field by a factor of $r$, its radial part resembles the 1-dimensional wave equation with a potential, $V(r)$. At large $r$, $V(r) \approx -(\omega^2 - \mu^2) $ where $\omega$ is the frequency of the field and $\mu$ its mass. If $\omega^2 - \mu^2 > 0$ then the field  would be outgoing, i.e. $\exp(i\sqrt{\omega^2-\mu^2} r_*)$, however, if  $\omega^2 - \mu^2 \le 0$ then the  field would fall off at large $r$, i.e. $ \exp(-\sqrt{\mu^2-\omega^2} r_*)$. Now, if the super-radiant condition is satisfied then the field is amplified, but cannot be radiated away (assuming $\omega \le \mu$) thus causing the instability. 
    
In this article, we will study this instability arising from the super-radiant scattering of a massive scalar field off a Kerr black hole. We perform this study using black hole perturbation theory in the time-domain i.e. using a numerical Teukolsky code \cite{pablo}. We present temporal profiles of the unstable scalar field and also estimate the growth rate of the instability. We compare our time-domain based results with past results based on frequency-domain calculations. 

\section{Black hole bomb}

In this section, we will demonstrate the unbounded growth of the amplitude of a massless scalar and also an electromagnetic wave packet in the context of super-radiant scattering off a Kerr black hole which is surrounded by perfect mirrors. We will solve the Teukolsky equation to study the behavior of the fields in the time-domain. This equation in Boyer-Lindquist coordinates, appears below. 
\begin{eqnarray}
&&
\Biggr\{\left[a^2\sin^2\theta-\frac{(r^2 + a^2)^2}{\Delta}\right]
\partial_{tt}-
\frac{4 M a r}{\Delta}\partial_{t\varphi}
-2s\left[r+ia\cos\theta-\frac{M(r^2-a^2)}{\Delta}\right]\partial_t
\nonumber\\
&&+\,\Delta^{-s}\partial_r\left(\Delta^{s+1}\partial_r\right)
+\frac{1}{\sin\theta}\partial_\theta\left(\sin\theta\partial_\theta\right)
+\left[\frac{1}{\sin^2\theta}-\frac{a^2}{\Delta}\right]
\partial_{\varphi\varphi} \nonumber\\
&&+\, 2s \left[\frac{a (r-M)}{\Delta} + \frac{i \cos\theta}{\sin^2\theta}
\right] \partial_\varphi
-\left(s^2 \cot^2\theta -s \right)\Biggr\}\Psi=0,
\end{eqnarray}
where $M$ is the mass of the black hole, $a$ its Kerr parameter,  $\Sigma\equiv r^2+a^2\cos^2\theta$,  $\Delta\equiv r^2-2Mr+a^2$ and $s$ is the spin-weight of the field. We use a $2+1$ dimensional Teukolsky evolution code \cite{pablo} in which one performs the numerical integration of the resulting equation upon decomposing the above equation into azimuthal angular modes. The boundary conditions used by the numerical code correspond to ``ingoing'' fields at the horizon and zero (``perfect mirror'') at the outer-boundary.  This code exhibits convergence as has been indicated before in the literature \cite{pablo}. 

Next, we will discuss the condition of so-called, super-radiance. If we consider a wave of frequency $\omega$ at the horizon, its spatial dependence is of the form, $\exp(\pm i(\omega-m\omega_{+}) r_*)$, where $m$ is the azimuthal mode of the wave and $\omega_{+}$ is the angular velocity of the event horizon of the hole, given by $\omega_{+} = {a \over {2Mr_{+}}}$ (where $r_{+}=M+\sqrt{M^{2}-a^{2}}$). For the physical boundary condition of ``ingoing'' wave, we insist on a negative group velocity, and therefore choose the fields to have a form,  $\exp(-i(\omega-m\omega_{+}) r_*)$. However, note that if $\omega < m\omega_{+}$ then the phase velocity will be positive! This is the super-radiance condition and it implies that energy and angular momentum are being extracted from the hole. 

Imposing a condition on frequency is difficult when attempting a time-domain evolution, however we use the idea of an ``almost monochromatic'' wave-packet as presented in \cite{pablo}. Specifically, we assume that the initial pulse ($\ell=m=1$) is centered far away from the hole at $r_{*}=r_{0}$ and that the modulation frequency is $\omega$. Therefore, 
\begin{equation}
\Psi (t=0) =  e^{-(r_{*}-r_{0})^{2}/b^{2}-i\omega(r_{*}-r_{0})}  \sin\theta
\label{id}
\end{equation}
In our evolutions, we chose $r_{0}=80M$, $\omega=0.25/M$ and a black hole background with $a=0.9999M$. Note that the $\omega$ chosen is in the super-radiant regime. The wave packet should have a wide spatial spread, so as to minimize the frequency overlap with the non-super-radiant regime. Therefore we chose, $b=50M$. The initial time derivative of the field was chosen so as to make this packet ``ingoing''. The numerical code related parameters, i.e. location of the inner boundary and the location of the outer boundary, where chosen to be $-40M$ and $160M$ respectively. Typical grid resolution used for the radial coordinate was $M/10$, while it was $\pi/32$ for the polar angle coordinate.    

The results of our computation appear in Figures 1 and Figures 2 below. The plotted waveforms were extracted at $r_{*} = -20M$ and $\theta=\pi/2$. It clear from the results that when $\omega$ is chosen to be in the super-radiance regime, a black hole bomb is created. The amplitude of the fields clearly grows without bound. 

\begin{figure*}
\includegraphics{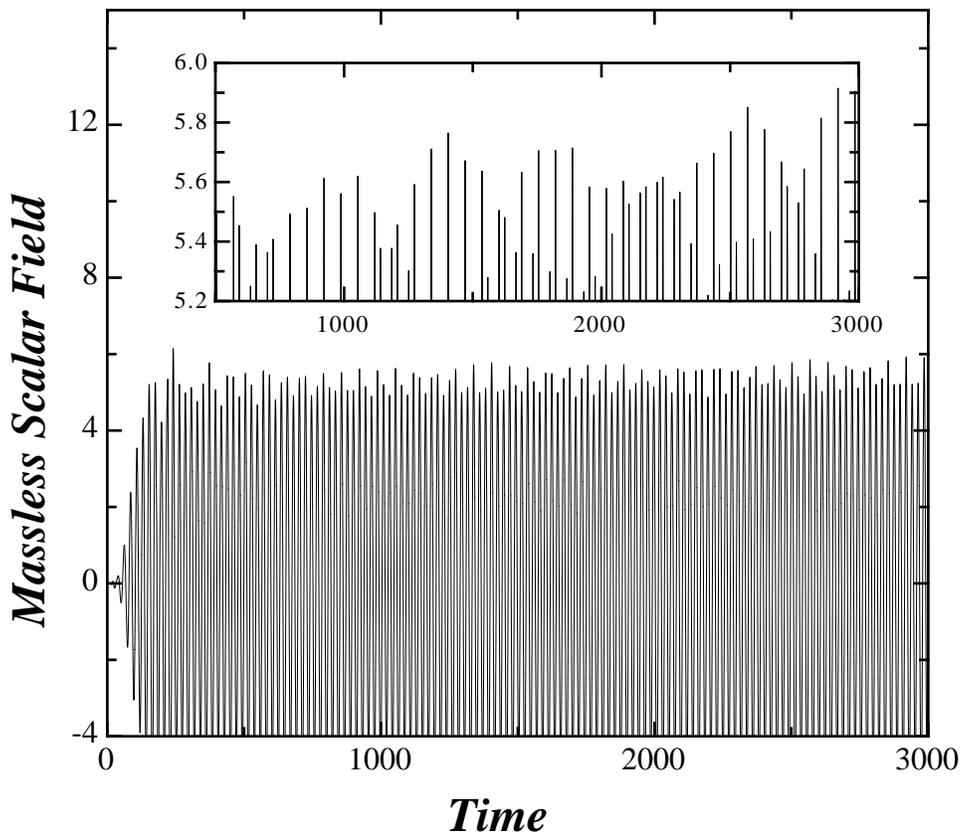}
\label{mirror1}
\caption{Black hole bomb: Massless scalar field sampled at $r_{*} = -20M$ in Kerr spacetime with $a/M=0.9999$ surrounded by perfect mirrors. The field grows without bound upon successive reflections. All quantities in units of black hole mass, $M$.}
\end{figure*}

\begin{figure*}
\includegraphics{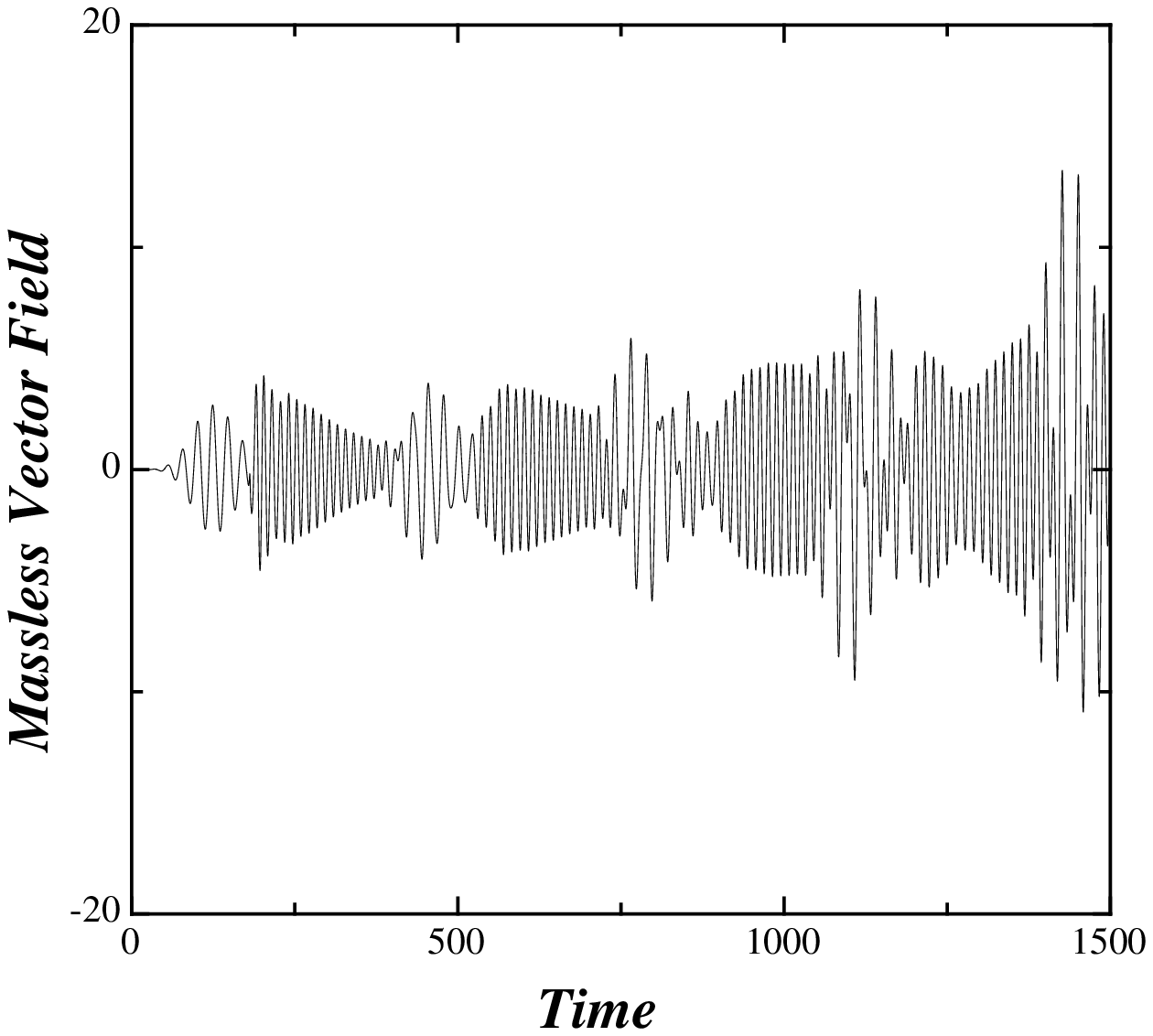}
\label{mirror2}
\caption{Black hole bomb: Electromagnetic field sampled at $r_{*} = -20M$ in Kerr spacetime with $a/M=0.9999$ surrounded by perfect mirrors. The field grows without bound upon successive reflections. All quantities in units of black hole mass, $M$.}
\end{figure*}

\section{Massive scalar field instability}

In this section, we will demonstrate the unbounded growth of the amplitude of a massive scalar wave packet in the context of super-radiant scattering off a Kerr black hole. We use the same basic formalism as in the previous section. We simply add a mass term to the Teukolsky equation of the form $\mu^{2}\Sigma\Psi$ and use the same numerical code with this minor modification. We conducted a detailed convergence test on this modified code. In summary, the code exhibits second-order convergence to about $t=500M$, and then its convergence order decreases slowly and settles to a bit above first-order at very late times  \footnote{Note that a time-domain study of late-time tails of massive scalar fields in black hole spacetimes has been performed before using a Teukolsky code very similar to this one (just in different coordinates) \cite{massive}. That work demonstrated that the late-time tail of massive scalars with axisymmetry, in both Schwarzchild and Kerr spacetimes is the same, i.e. $t^{-5/6}$. }. 

To witness the instability in this case, in addition to satisfying the super-radiance condition, $\omega<\omega_{+}$, the frequency of the wave-packet also needs to satisfy the condition, $\omega  \le \mu$. As argued in the introduction section of this article, this is necessary because it ensures that the field cannot be radiated away. The condition $\omega \le \mu$ forces the fields at large $r$ to decay exponentially with $r$, thereby ``trapping'' it in the system i.e. the mass plays the role of a natural mirror. The field grows because of super-radiance, but is a bound-state, and that is the reason it becomes unstable. 

For the initial pulse, again we chose data of the form indicated in equation \ref{id}. This time, we chose $r_{0}=120M$, $\omega=0.25/M$ and a black hole background with $a=0.9999M$. The mass of the scalar field is chosen to be $\mu=0.25/M$, and with these choices the conditions for instability are satisfied. We again chose the wave-packet to have a wide spread, $b=50M$. The location of the inner boundary and the location of the outer boundary, where chosen to be $-40M$ and $1460M$ respectively. The outer boundary condition was chosen to be that for ``outgoing'' fields. Note that as much as possible, we use frequencies that are integer multiples of each other to avoid excessive mixing or ``beats'' because that makes studying the late-time behavior of the fields difficult.  

The results from this evolution ($\ell=m=1$) appear in Figure 3. The data has been extracted from the spatial location,   $r_{*} = 20M$ and $\theta=\pi/2$. The plot clearly demonstrates the presence of an instability although its growth is very slow. Because of the slow growth, it is difficult to judge the functional form of the growth (exponential, power-law, etc.) however, one can obtain the ``local'' linear growth rate by computing the slope of the envelope of the depicted growing oscillations. We estimate that this rate to be about $2 \times 10^{-5}/M$ from our results. If we assume that the growth is actually exponential, this would translate to an e-folding time of about $5 \times 10^{4}M$. From the perspective of the size of our computation (recall, we are evolving a $2+1$ dimensional code) this e-folding time is very large. It would take a non-trivial effort to make suitable modifications to the numerical method used by our code to accurately run that long. It would also require tremendous amount of computational power. Therefore, we leave that open for sometime in the future. We may also attempt a survey of the parameter space, in order to understand the dependence of the growth rate on the parameters, although that may be best done in the frequency-domain (see, for example \cite{old}). Also, it should be noted that the results we have appear to be two orders of magnitude larger than the approximate growth rates that have been computed from the frequency-domain based computations. In particular, the e-folding time can be computed \cite{detweiler}, using $(a/M)^{-1}24(\mu M)^{-8}\mu^{-1}$ which turns out to be $6\times 10^{6}M$ for the parameters relevant to our case. We attribute this apparent discrepancy to the use of the approximation, $M\mu << 1$ used in deriving above expression  which does not quite apply to our choice of parameters. Unfortunately, we are unable to make a comparison with a significantly lower value of $M\mu$ because that would increase the e-folding time tremendously, and we would be unable to make an estimate of such a small growth rate because of the relatively short time-scale of our evolutions.  
  
\begin{figure*}
\includegraphics{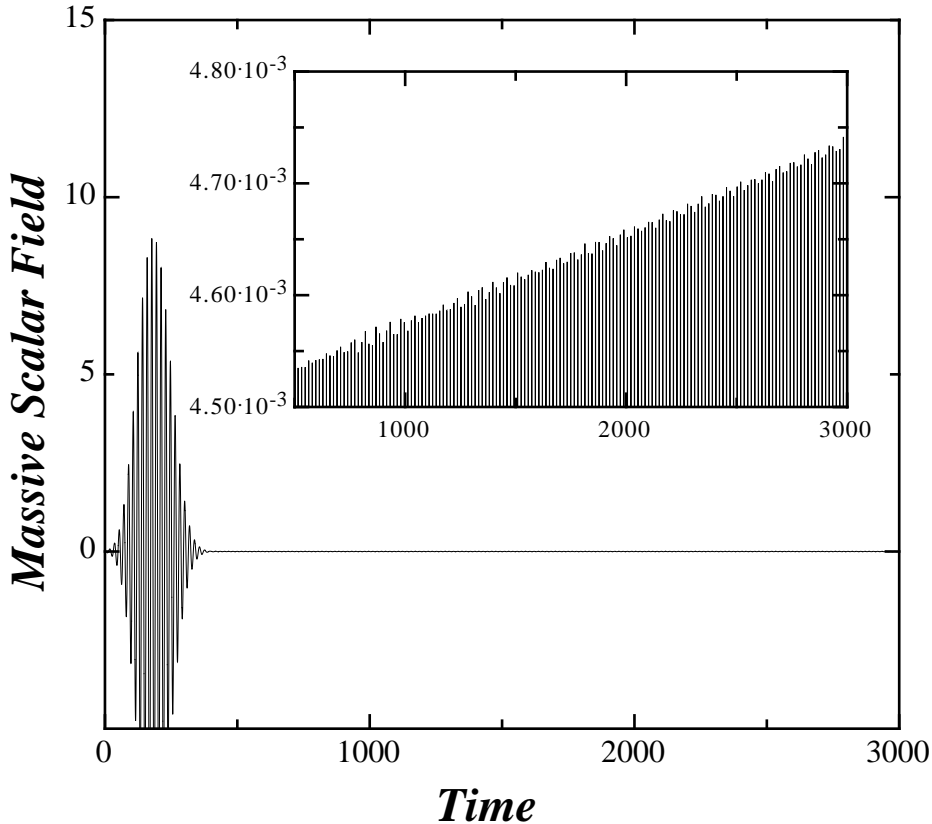}
\label{instability}
\caption{Instability: Massive scalar field ($\ell=m=1$) sampled at $r_{*} = 20M$ in Kerr spacetime with $a/M=0.9999$. The field grows without bound, demonstrating the existence of instability. All quantities in units of black hole mass, $M$.}
\end{figure*}

We also show results from the case in which the only change we make is changing the sign of the frequency of the initial wave packet i.e. set $\omega=-0.25/M$. This $\omega$ violates the super-radiance condition and therefore, we no longer expect to see an instability. And that is indeed what the Figure 4 indicates. 

\begin{figure*}
\includegraphics{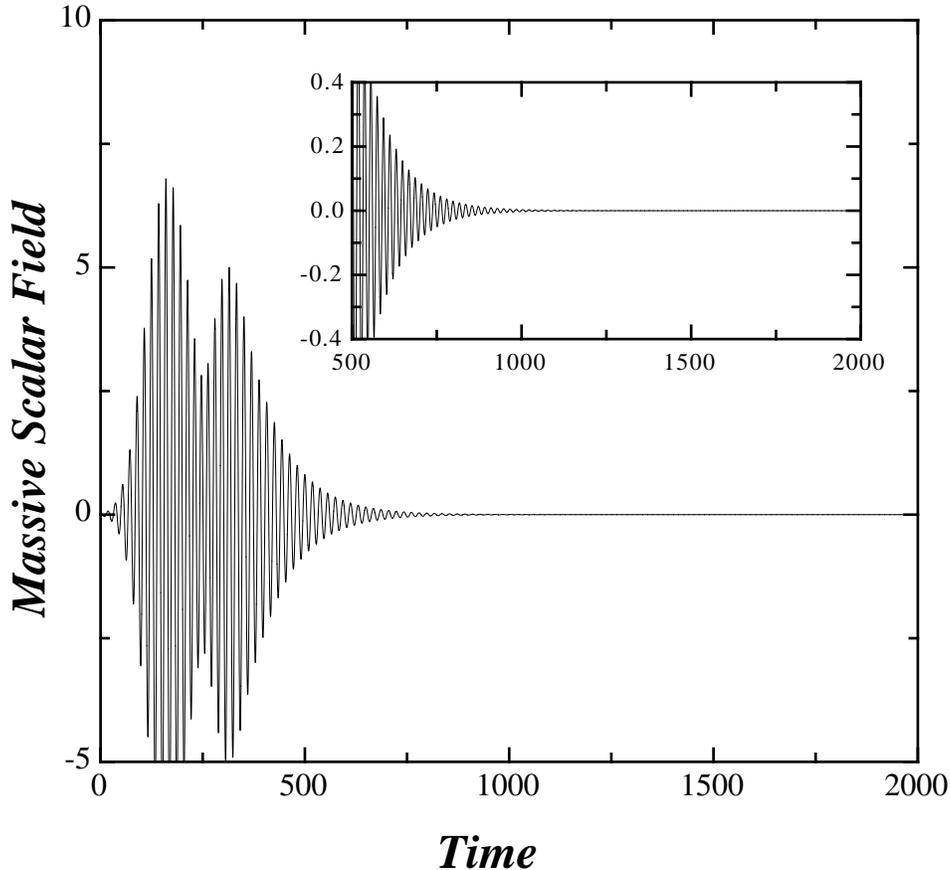}
\label{noinstability}
\caption{Massive scalar field ($\ell=m=1$) sampled at $r_{*} = 20M$ in Kerr spacetime with $a/M=0.9999$. Here the super-radiance condition is not satisfied. The field decays, demonstrating the absence of an instability. All quantities in units of black hole mass, $M$.}
\end{figure*}

In Figure 5 we show similar results from the evolution of the $\ell=m=2$ mode. We used $\omega=0.5/M$, $\mu=0.5/M$ and all the other parameters were left at the same values as before. The growth rate of the instability in this case, is about an order of magnitude lower than the $\ell=m=1$ case above. 

\begin{figure*}
\includegraphics{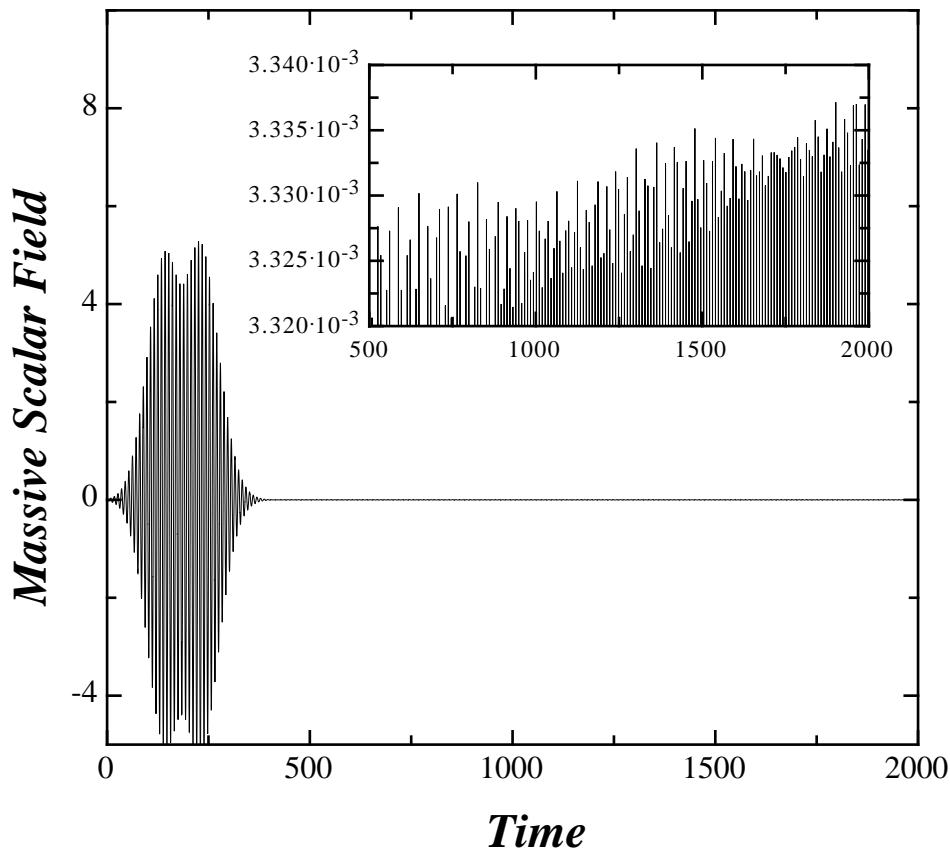}
\label{instability2}
\caption{Instability: Massive scalar field ($\ell=m=2$) sampled at $r_{*} = 20M$ in Kerr spacetime with $a/M=0.9999$. The field grows without bound, demonstrating the existence of instability. All quantities in units of black hole mass, $M$.}
\end{figure*}

\section{Conclusions}

Using a time-domain $2+1$ dimensional Teukolsky code we have been able to demonstrate that under conditions of super-radiance, a black hole bomb can be constructed using ``artificial'' or ``natural'' mirrors. We demonstrated that massless scalar and electromagnetic wave packets grow without bound if they satisfy the condition of super-radiance, as they scatter off a Kerr hole that is surrounded by mirrors. We also demonstrated the result of using a natural mirror, i.e. the mass of a scalar field to the achieve the same effect. Unfortunately, since the growth rate of the instability is very low, it was only possible to make a very crude estimate of the e-folding time. However, even with the results we obtained, it seems that the growth rate of the instability could be two orders of magnitude larger than what was known before, which could have some interesting astrophysical implications. 

\section{Acknowledgments}
GK thanks Jorge Pullin for suggesting this project. GK also thanks Steven Detweiler and Lior Burko for comments, discussion and advice regarding this work. We thank the University of Massachusetts at Dartmouth for support and also acknowledge research support from the UMass Healey Endowment and Glaser Trust (Joseph Hanshe, Trustee).


\begin{references}

\bibitem{detweiler} S. Detweiler, Phys. Rev. D {\bf 22}, (1980) 2323

\bibitem{old} V. Cardoso, O. J. C. Dias, J. P. S. Lemos, S. Yoshida, Phys. Rev. D {\bf 70} (2004) 044039 \\
H. Furuhashi, Y. Nambu, gr-qc/0402037

\bibitem{pablo} W. Krivan, P. Laguna, P. Papadopoulos, N. Andersson, Phys. Rev. D {\bf 56} (1997) 3395

\bibitem{pablo-scalar} N. Andersson, P. Laguna, P. Papadopoulos, Phys. Rev. D {\bf 58} (1998) 087503 \\
N. Andersson, K. Glampedakis, Phys. Rev. Lett. {\bf 84} (2000) 4537

\bibitem{massive}  L. M. Burko, G. Khanna, Phys. Rev. D {\bf 70} (2004) 044018

\bibitem{rapid}  L. M. Burko, G. Khanna, Phys. Rev. D {\bf 67} (2003) 081502
 
\bibitem{teuk}  S. A. Teukolsky, Astrophys. J., {\bf 185}, (1973) 635

\bibitem{press} W. H. Press, S. A. Teukolsky, Nature, {\bf 238}, (1972) 211

\bibitem{damour} T. Damour, N. Deruelle, R. Ruffini, Lett. Nuovo Cimento {\bf 15}, (1976) 257






\end{references}
\end{document}